\documentclass{PoS}

\title{Perspectives for Top quark physics at High-Luminosity LHC}

\ShortTitle{Perspectives for top physics at HL-LHC}

\author{Michele Selvaggi
\\
Centre for Cosmology, Particle Physics and Phenomenology (CP3)\\
Universit\'e catholique de Louvain, Chemin du Cyclotron 2, B-1348 Louvain-la-Neuve, Belgium\\
        E-mail: \email{michele.selvaggi@uclouvain.be}}

\abstract{The High-Luminosity LHC is expected to provide 3 $ab^{-1}$ of integrated luminosity. As a result billions of events containing top quarks will be detected at the CMS and ATLAS experiments, allowing for precise measurements of the top quark properties. The experimental challenges that will be faced in a high luminosity environment, with a special focus on top quark related observables are examined. We discuss prospects for measuring top quark anomalous couplings at the HL-LHC. Projections for detecting flavor changing neutral currents involving top quarks are also reviewed.} 

\FullConference{8th International Workshop on Top Quark Physics, TOP2015\\
		14-18 September, 2015\\
		Ischia, Italy}

\begin{document}

\section{Experimental challenges}

The program of the High-Luminosity (HL-LHC) is to deliver over the course of 10 years 3000 $fb^{-1}$ of data with an instantaneous luminosity peaking at $5 \times 10^{34} cm^2 s^{-1}$ starting from 2025. Such unprecedented LHC performance comes with several experimental challenges and puts serious constraints on the ATLAS and CMS experiments to maintain present performance in such an extreme environment~\cite{cms_upgrade,atlas_upgrade} . The typically higher event rates compared to Run 2 will be addressed by both experiments by implementing specific triggers using track reconstruction. At such luminosities the amount of p-p collisions per bunch crossing (pile-up) is expected to peak at 150 - 200 simultaneous interactions. The high levels of radiation endured during Run 2, together with the extreme pile-up conditions will require highly granular pixel detectors.   

\begin{figure}[bh!]
\centering
\begin{minipage}{.33\textwidth}
  \centering
  \includegraphics[width=0.90\linewidth]{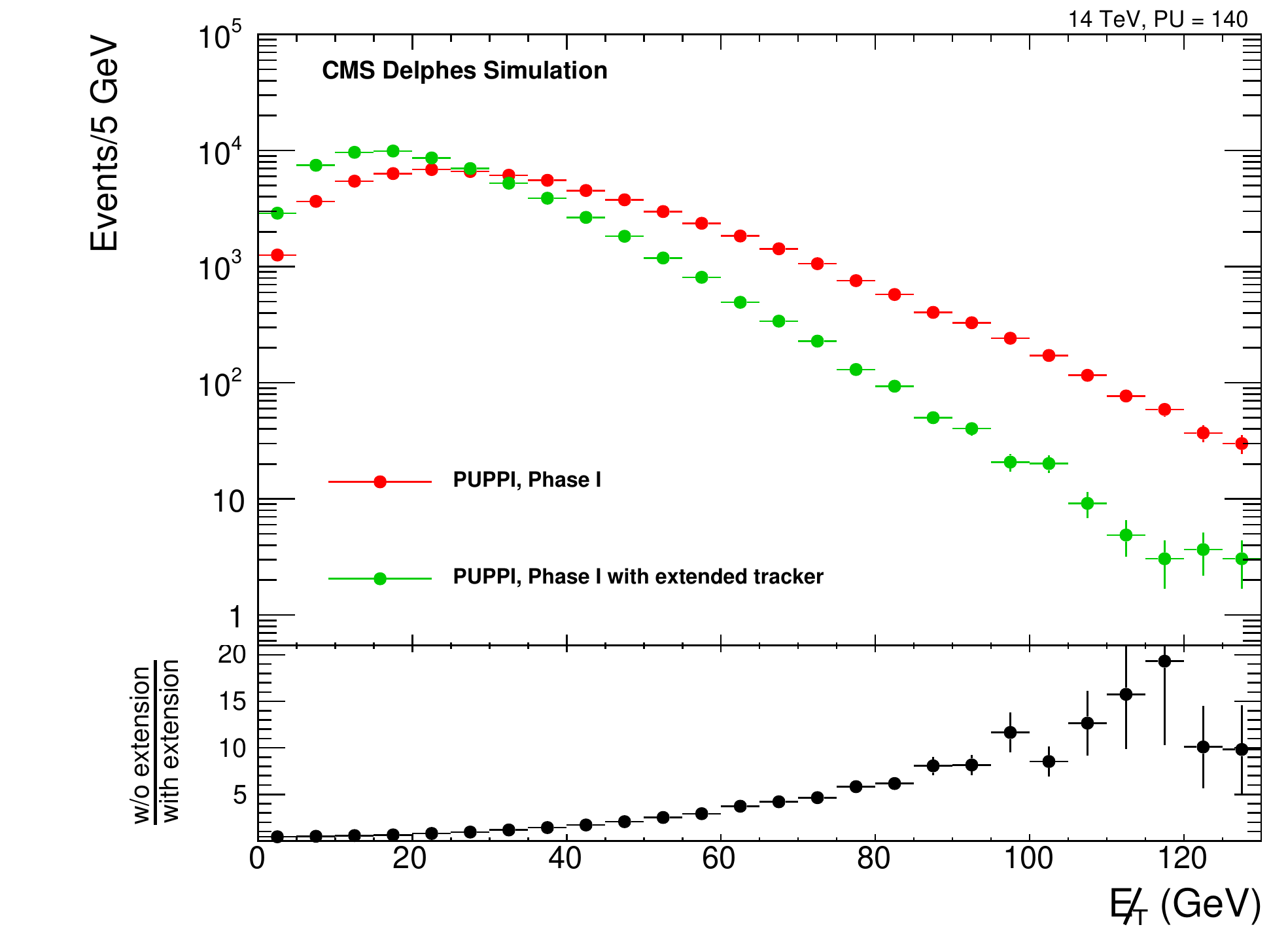}
  \label{fig:1a}
\end{minipage}%
\begin{minipage}{.33\textwidth}
  \centering
  \includegraphics[width=0.90\linewidth]{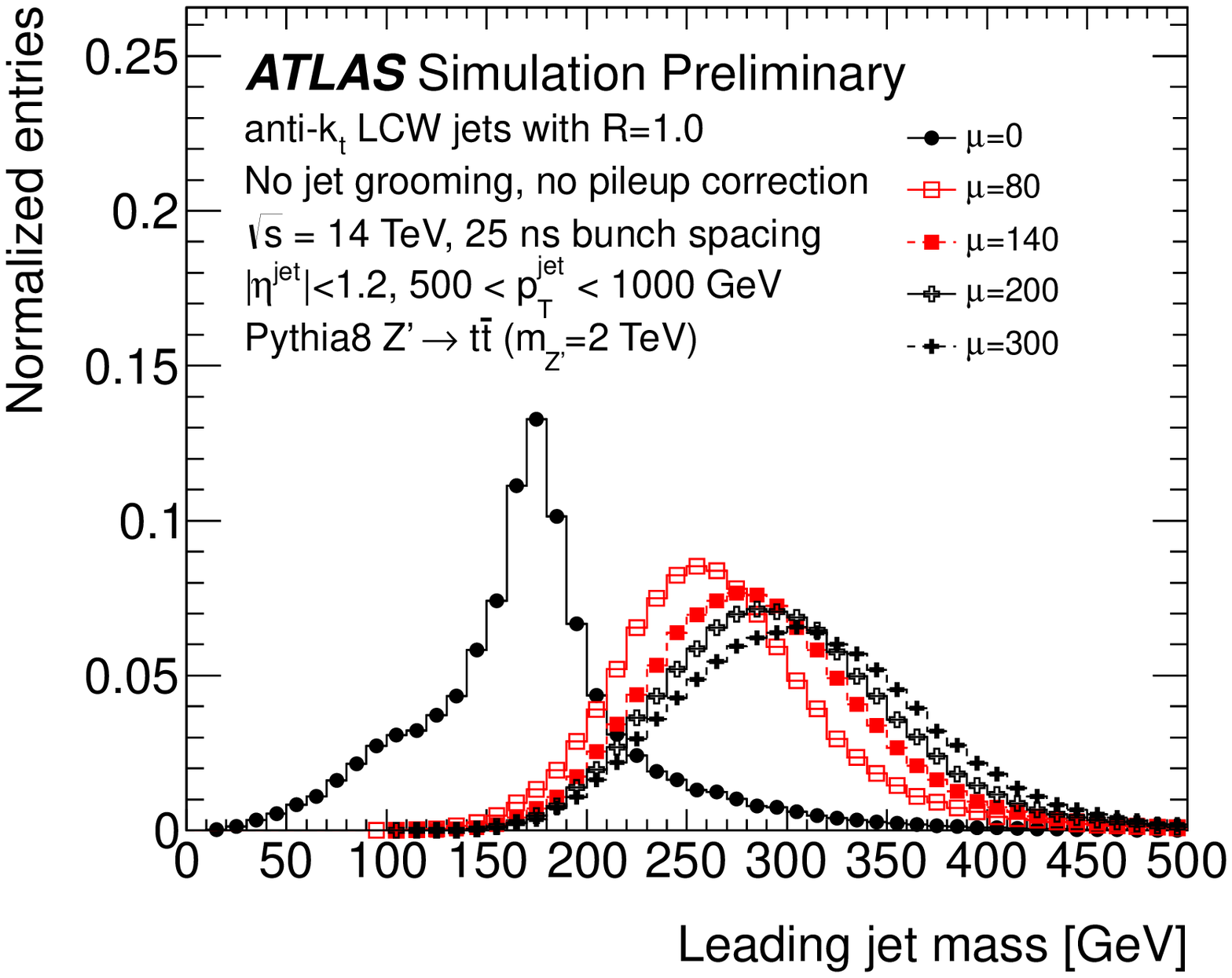}
  \label{fig:1b}
\end{minipage}%
\begin{minipage}{.33\textwidth}
  \centering
  \includegraphics[width=0.90\linewidth]{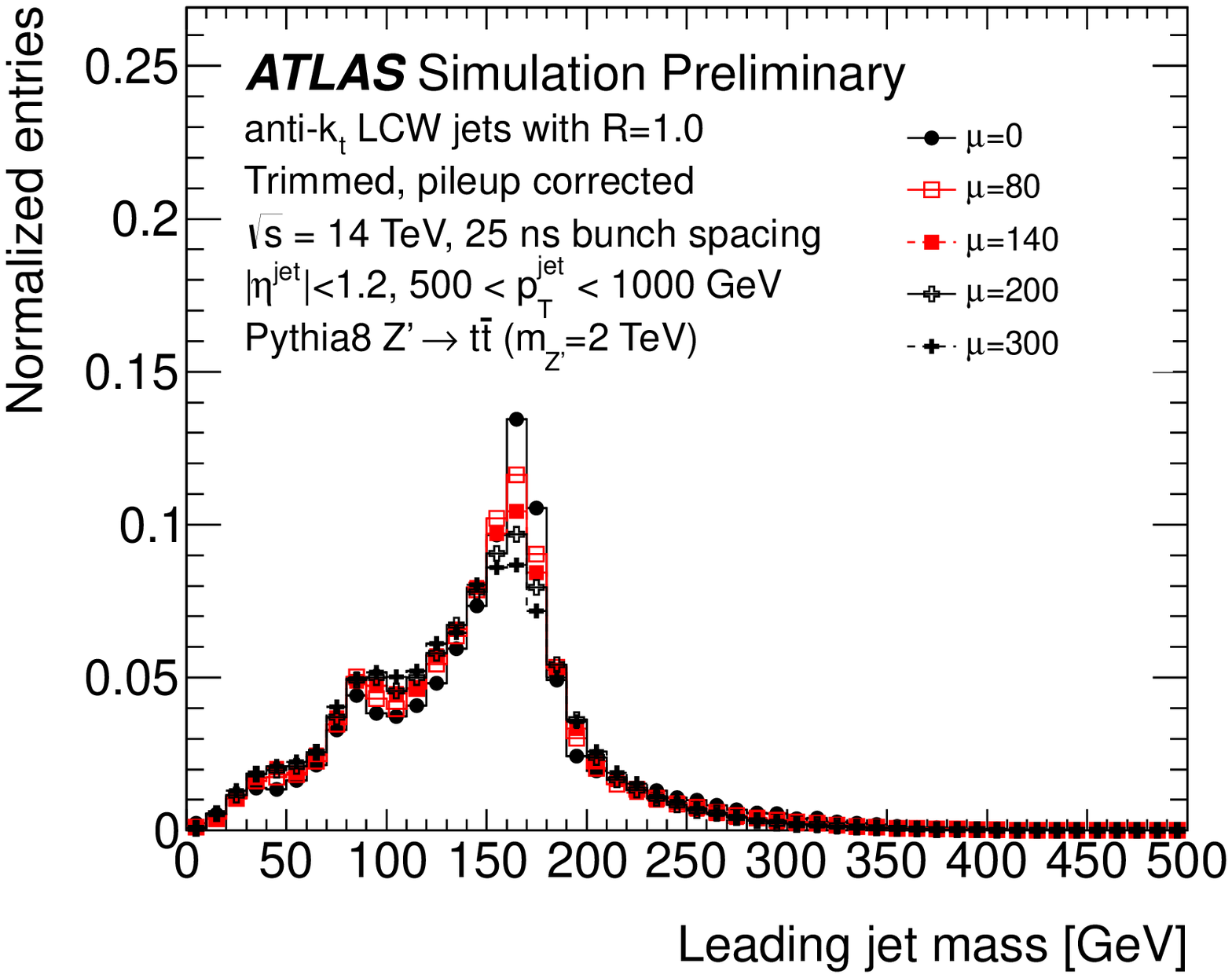}
  \label{fig:1c}
\end{minipage}%
\begin{center}
\caption{\label{fig:run}\emph{Left: Comparison of the $E_T^{miss}$ resolution with the present CMS tracker and with an hypothetical extended tracker up to $|\eta| = 4$~\cite{cms_upgrade,deFavereau:2013fsa}; Center: Ungroomed jet mass; Right: Groomed jet mass ~\cite{atlas_web}.}}
\end{center}
\end{figure}

The top quark decays almost exclusively into a b-quark and either to two light jets or to a lepton and a neutrino. The presence of a top quark can be therefore inferred by the presence of a b-jet plus a lepton and missing transverse energy ($E_T^{miss}$). Most tops quark pairs being produced at low momentum (at threshold) at the LHC, its decay products will also carry low momentum ($p_T\approx 30$ GeV). At low momentum the performance of $E_T^{miss}$ reconstruction and b-jet identification as well as jet energy resolution can be highly degraded by extreme pile-up conditions. Such effects can potentially result in poor top reconstruction performance. In addition to high granularity sub-detectors, extended detector acceptance and advanced pile-up substraction algorithms~\cite{Bertolini:2014bba} can help in recovering good performance. The improvement in the $E_T^{miss}$ resolution using an extended tracker up to $|\eta| < 4$ is illustrated in figure~\ref{fig:1a} (left plot).    

The top pair production cross-section at the LHC being so large at the LHC, millions of top quarks with a large transverse boost ($p_T > 500$ GeV) will also be produced and studied. The decay products of a top quark with $p_T = 500$ GeV are typically separated by $\Delta R \approx 2 m_t / p_T \approx 0.8$, which is comparable to the typical size of jets reconstructed at the LHC. Observables that measure how the energy deposits are distributed within such characteristic jets~\cite{Thaler:2010tr}, or the jet mass distribution can be used in order to discriminate between boosted "top jets" and background jets originating from light partons. However such observables can also be largely sensitive to soft contamination in the event, both diffuse (initial state radiation or pile-up), and local (final state radiation, especially from the top itself) as shown in figure~\ref{fig:1b} (center). Such degradation can be cured by applying \emph{jet grooming} techniques~\cite{Krohn:2009th} as shown in figure~\ref{fig:1c} (right plot). 

\section{Measurements at high luminosity LHC}

The top quark carries colour, yet it is the only quark that decays before it hadronize, giving us a great opportunity to study its bare properties. For instance, precisely measuring its mass is crucial for understanding the stability of the vacuum of our universe~\cite{stab}. Spin correlation among its decay products and forward-backward asymmetries can be sensible to new physics~\cite{Jung:2013vpa,Frampton:2009rk}. Using an effective field theory approach (EFT)~\cite{Grzadkowski:2010es}, we can set limits on the scale of new physics by precisely measuring the coupling of the top with other standard model particles. One can indirectly probe new physics by studying rare top decays such as those involving flavour changing neutral currents (FCNC's). Finally, the presence of one or several top quarks in the final state can result from the direct production of heavy non-SM states. In this note we will focus on measurements of the couplings and FCNC's.

\subsection{Couplings}

By complying with Lorentz invariance, one can extend the top-gluon QCD coupling in the SM by adding contributions that lead to chromo-electric and chromo-magnetic dipole moments~\cite{AguilarSaavedra:2008zc}:

\begin{equation}
\mathcal{L}_{gtt} = -g_s \bar{t}\gamma^\mu\frac{\lambda_a}{2} t G^a_{\mu} + \frac{g_s}{m_t}\bar{t}\sigma^{\mu\nu}(d_V + i d_A \gamma_5)\frac{\lambda_a}{2} t \mathcal{G}^a_{\mu\nu}
\end{equation} 

The parameters $d_V$ and $d_A$ are vanishing at leading order in the SM. In principle the total top pair cross section is directly sensitive to the \emph{gtt} coupling given that more than 90\% of the $t\bar{t}$ pair are produced via gluon-fusion at LHC. However one can enhance the contribution from the chromo-moments by measuring events with large invariant mass, given the explicit dependence on the internal gluon momentum in the interaction.  This measurement, proposed in ref.~\cite{Aguilar-Saavedra:2014iga} can possibly lead to the constrain $|d_{A,V}| < 0.02$ at the HL-LHC.

\begin{figure}[b]
\centering
\begin{minipage}{.5\textwidth}
  \centering
  \includegraphics[width=0.75\linewidth]{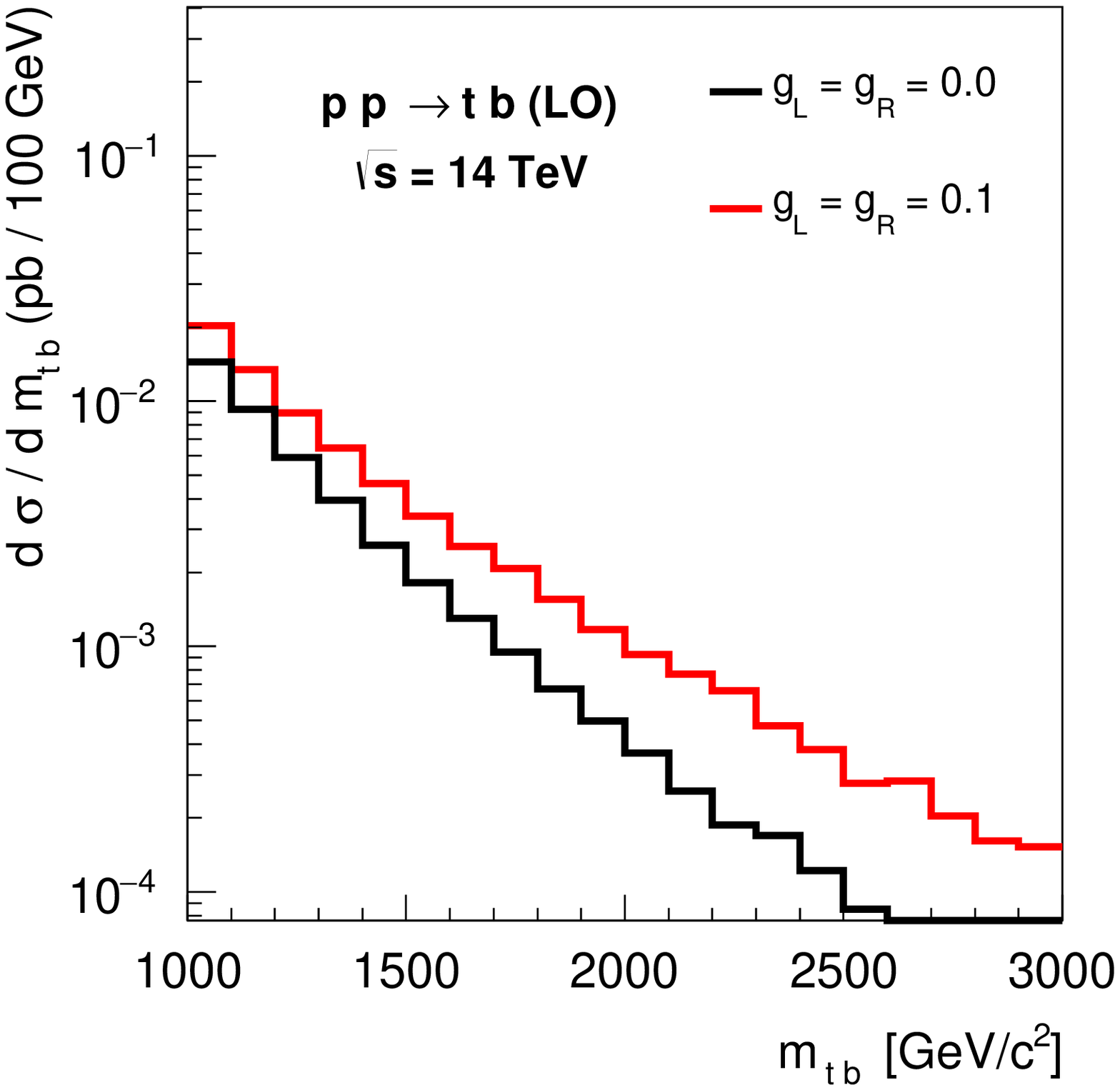}
  \label{fig:2a}
\end{minipage}%
\begin{minipage}{.5\textwidth}
  \centering
  \includegraphics[width=0.95\linewidth]{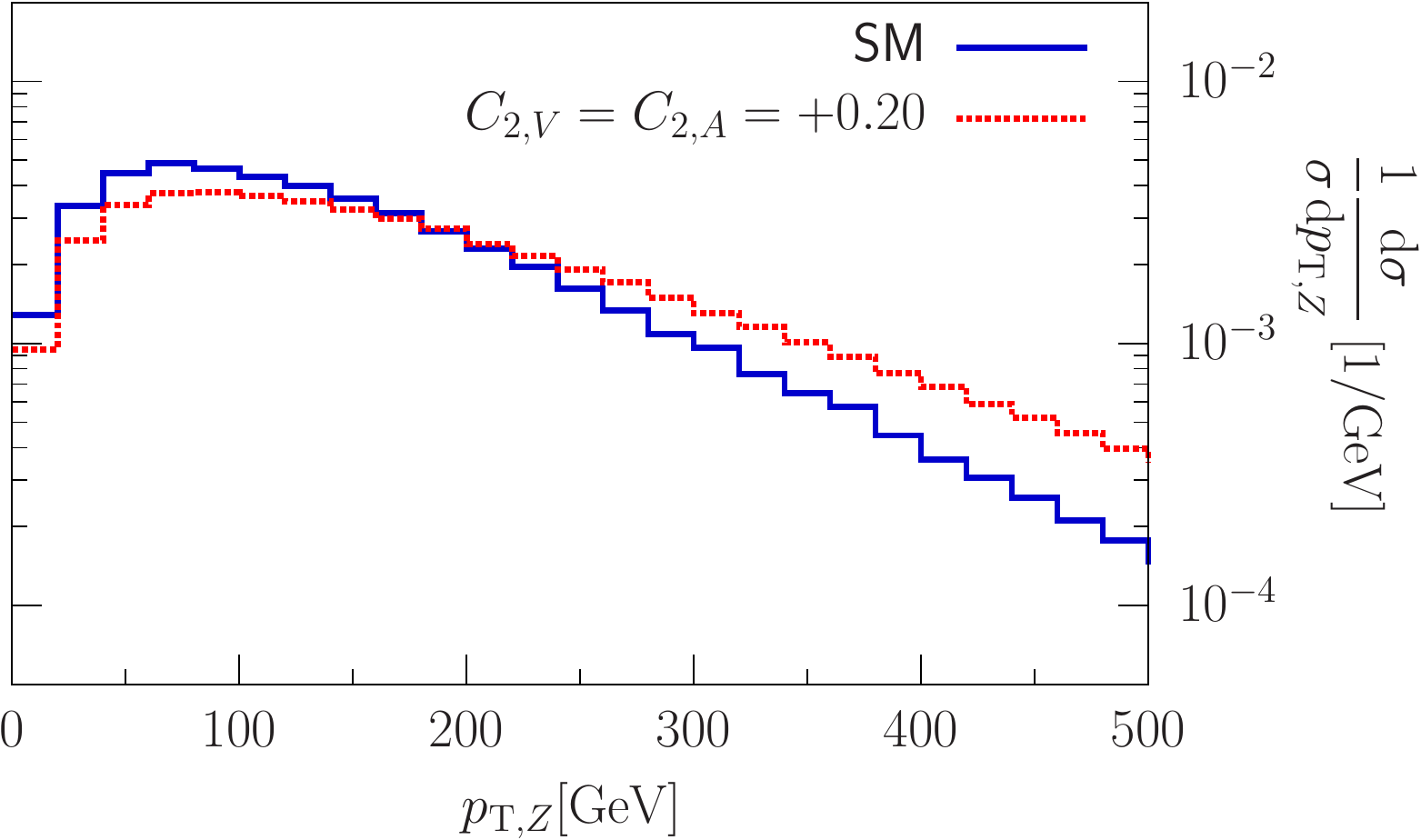}
  \label{fig:2b}
\end{minipage}%
\begin{center}
\caption{\label{fig:run}\emph{Left: Invariant mass distribution of the (t,b)  final state in single top s-channel production in the SM ($g_L = g_R = 0$) and with anomalous couplings ($g_L = g_R = 0.1$); Right: Transverse momentum distribution of Z in $ttZ$ production in the SM $C^Z_{2,A} = C^Z_{2,V} = 0$ and with anomalous couplings $C^Z_{2,A} = C^Z_{2,V} = 0.2$~\cite{Rontsch:2014cca}.}}
\end{center}
\end{figure}

The $tWb$ interaction term can be written as:

\begin{equation}
\mathcal{L}_{tWb} = -\frac{g}{\sqrt{2}} \bar{b}\gamma^\mu(V_LP_L + V_RP_R) t W^-_{\mu} -\frac{g}{\sqrt{2}} \bar{b}\frac{i\sigma^{\mu\nu}q_\nu}{M_W}(g_LP_L + g_RP_R) t W^-_{\mu} + h.c. 
\end{equation} 

where $V_L = V_{tb}$ and $V_R = g_L = g_R = 0$ at tree level. The anomalous tensor-like coupling can be directly probed by measuring single top s-channel production at large momentum transfer as shown in figure~\ref{fig:2a} (left plot). The final state involves a $t$ and a $b$ quark at large $p_T$. A naive  analysis performed at leading order with {\sc MadGraph5\_aMC@NLO}~\cite{Alwall:2014hca} has shown that measuring $\sigma(m_{tb}> 2  \mathrm{TeV})$ can lead to $|g_{A,V}| < 0.10$ with 3 $ab^{-1}$. This result can probably be improved by including as a possible signal single top with associated W boson production in the boosted region.

Similarly, the $ttZ$ coupling can be generalized to:

\begin{equation}
\mathcal{L}_{ttZ} = e \bar{t}\gamma^\mu(C^Z_{1,V} + \gamma_5 C^Z_{1,A}) t Z_{\mu} + e \bar{t}\frac{i\sigma^{\mu\nu}q_\nu}{M_Z}(C^Z_{2,V} + i\gamma_5 C^Z_{2,A}) t Z_{\mu} + h.c. 
\end{equation} 

Anomalous couplings $C^Z_{2,A}$ and $C^Z_{2,V}$ can be probed by studying \emph{ttZ} production. Despite the small cross-section ($\approx$ 1 pb at 14 TeV), it has been shown that in the three lepton channel one can constrain the tensor like coupling to $|C^Z_{2,A}|,|C^Z_{2,V}| < 0.08$ by exploiting the potential enhancement of the cross section at large momenta (see figure~\ref{fig:2b}, right plot) and the difference in shape of the $\Delta \phi_{ll}$ distribution~\cite{Rontsch:2014cca}.

Finally, the top yukawa coupling, responsible for the large top mass,  can be measured through direct ttH production. Eventually at large integrated luminosity,  the sensitivity will most likely be dominated by the channels involving the smallest systematic uncertainties, namely $H\rightarrow\gamma\gamma$ and $H\rightarrow\tau\tau$.  Projections performed by the ATLAS~\cite{atlas_tth} and CMS collaboration~\cite{cms_upgrade} show similar sensitivity for precision on the top Yukawa coupling ($\Delta y_t/y_t \approx 10\%$). Other studies show that the $p_T^H$ distribution and the rapidity difference between the top decay products are particularly sensitive to the CP structure of the ttH coupling~\cite{Demartin:2014fia}.

\subsection{Flavour changing neutral currents}

Top induced FCNC's are largely suppressed by the GIM mechanism and by the large width of the $t\rightarrow Wb$ mode in the Standard model. As a result, any sizable evidence for FCNC's would be a compelling indication for new physics. Large FCNC's are predicted for instance by two Higgs doublet models (2HDM)~\cite{Luke:1993cy}, the MSSM~\cite{Cao:2007dk} as well as R-parity violating SUSY~\cite{Yang:1997dk} and Randall-Sundrum (RS) models~\cite{Agashe:2006wa}.   

Top FCNC's decays can be generically labelled as $t\rightarrow Xq$, where $X=g,\gamma, Z, H$ and $q = u, c$. Most estimates for top FCNC's HL-LHC sensitivity are obtained through naive extrapolations of statistical and systematic uncertainties at this day~\cite{Agashe:2013hma}. Genuine MC based analysis have only been performed in the context of $t\rightarrow Zq$ by CMS~\cite{CMS:2013zfa} and for $t\rightarrow cH$ by ATLAS~\cite{atlas_tch}. 

The most promising channel for $t\rightarrow Zq$ is the 3 leptons, 1 b-jet and missing energy final state, which is very clean, and mostly dominated by the ttV backgrounds. The analysis performed by CMS claims that the limit $BR(t\rightarrow Zq) < 10^{-4}$ for 3 $ab^{-1}$ is achievable, showing an improvement by a factor 10 from the limit $BR(t\rightarrow Zq) < 10^{-3}$ obtainted with 19 $fb^{-1}$ at 8TeV. The limit at 8 TeV has been recently improved by a factor 2~\cite{Chatrchyan:2013nwa}, it is therefore conceivable that $BR(t\rightarrow Zq) < 5\times 10^{-5}$ could be achieved with 3 $ab^{-1}$.    

For the $t\rightarrow c H$ study, ATLAS concentrated on the $H\rightarrow \gamma \gamma$ channel, assumed to be statitistics dominated at high integrated luminosity. The study has been performed both in the 4 jets final state, 2 photon (hadronic top) or 2 jets, lepton, di-photon final state. The final limit for 3 $ab^{-1}$ is $BR(t\rightarrow c H) < 1.5 \times 10^{-4}$. 

\begin{figure}[b]
\centering
\begin{minipage}{.5\textwidth}
  \centering
  \includegraphics[width=0.80\linewidth]{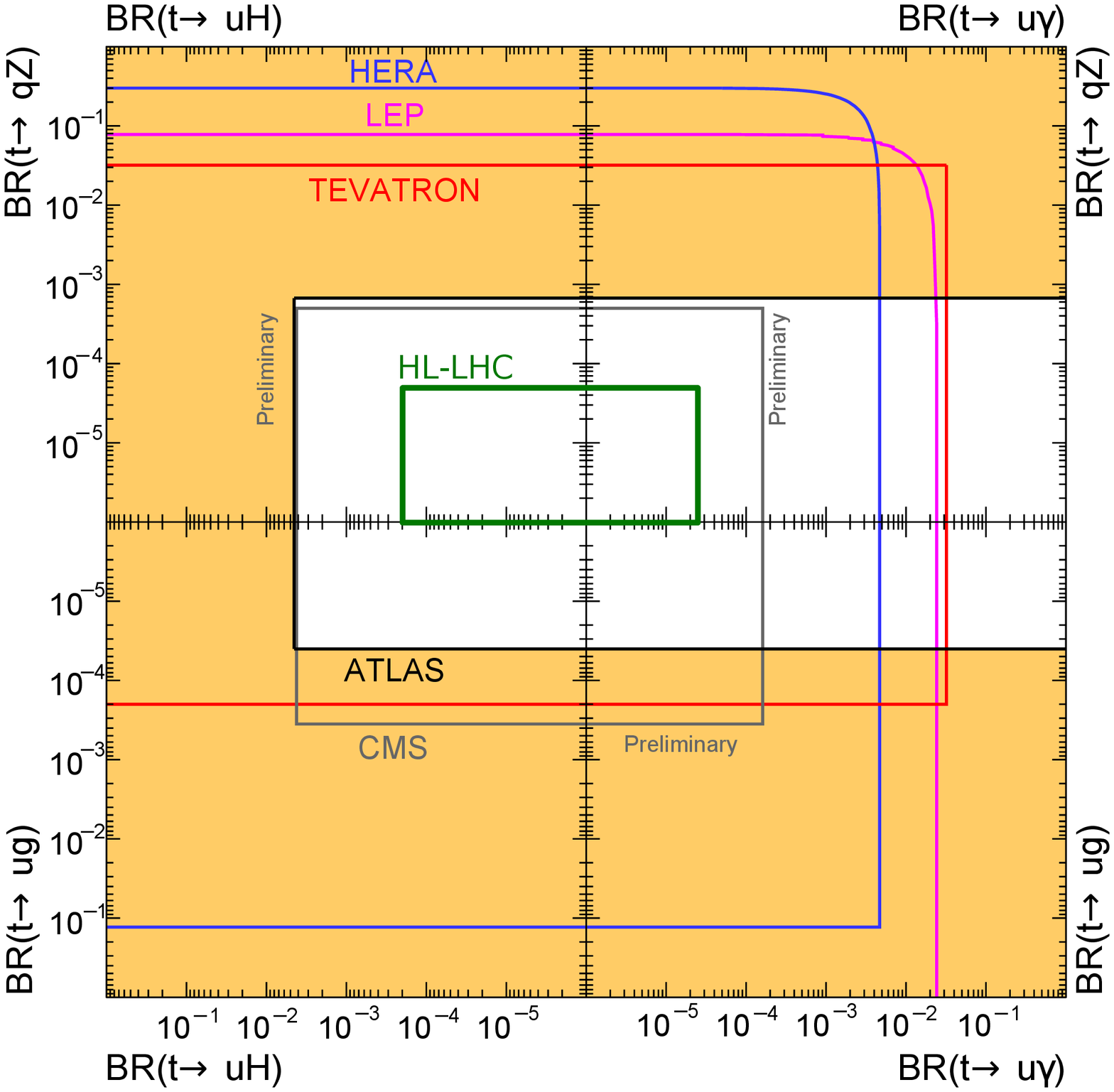}
  \label{fig:3a}
\end{minipage}%
\begin{minipage}{.5\textwidth}
  \centering
  \includegraphics[width=0.80\linewidth]{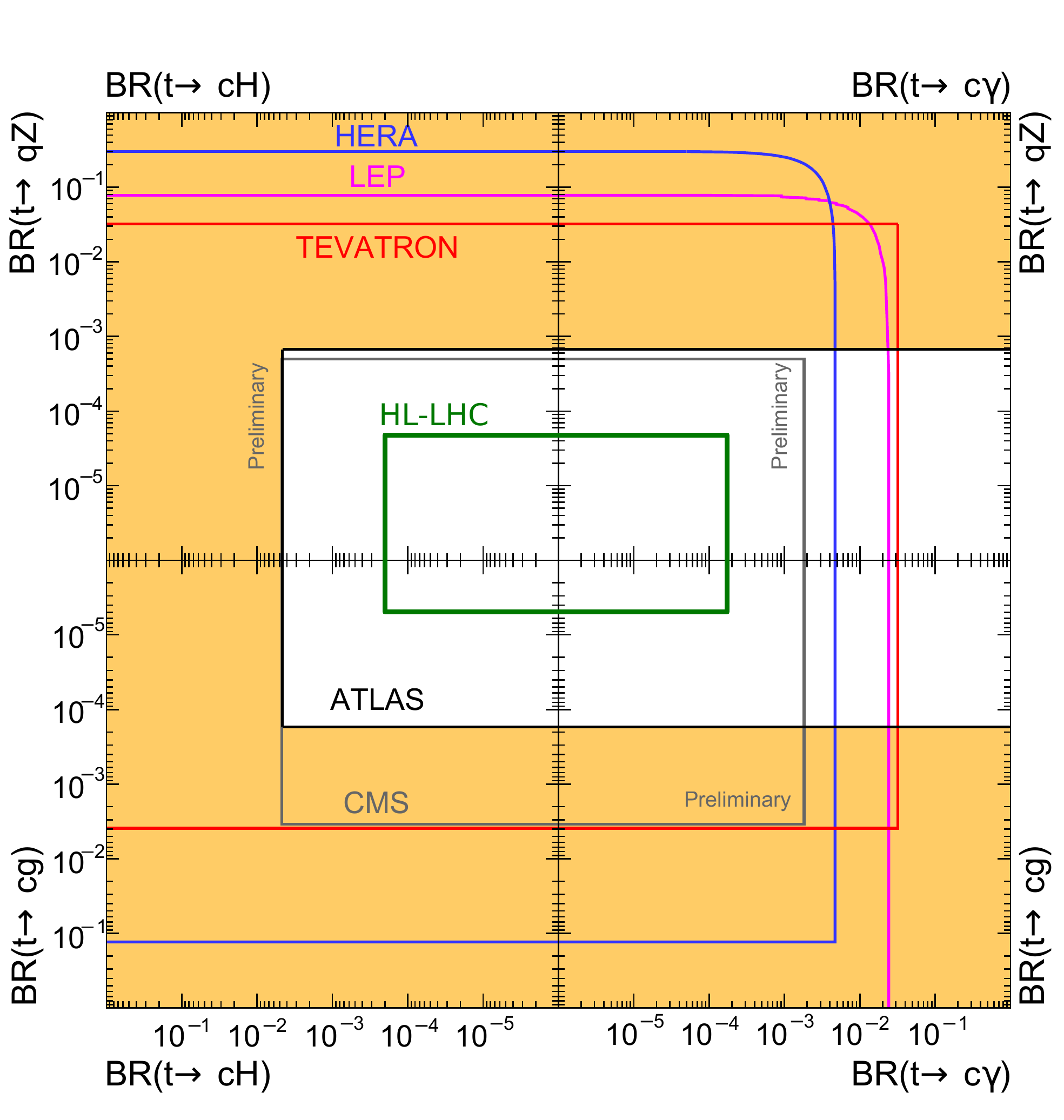}
  \label{fig:3b}
\end{minipage}%
\begin{center}
\caption{\label{fig:3}\emph{Summary for projections of top FCNC branching ratios for High luminosity LHC for $t\rightarrow Xq$ (left) and $t\rightarrow Xc (right)$, where $X=g,\gamma, Z, H$. The original figures, provided by the ATLAS collaboration can be found in~\cite{atlas_web}}. With the ecxeption of $t\rightarrow Zq$~\cite{CMS:2013zfa} and $t\rightarrow Hq$~\cite{atlas_tch}, the projections are taken from the SnowMass report~\cite{Agashe:2013hma}.}
\end{center}
\end{figure}

Most extrapolations for the remaining FCNC channels have been performed in the context of the SnowMass studies~\cite{Agashe:2013hma}. The obtained projections for the limits on $BR(t\rightarrow Xq)$ and $BR(t\rightarrow Xc)$ at the HL-LHC are summarized in figure~\ref{fig:3}. 

\section{Conclusion}

The high-luminosity LHC will be a machine operating at extreme conditions. In order to maintain today's performance, the capabilities of present detectors will have to be extended (track-based triggers, larger acceptance, high granularity). Billions of top quark will be produced. Most properties of the top quark could be measured at percent level precision, providing the opportunity to challenge the Standard Model as best as possible. 

\begin{acknowledgments}

We thank M. Mangano, A. Giammanco, F. Maltoni, F. Demartin and G. Durieux for helpful and stimulating discussions.

\end{acknowledgments}

\end{document}